\documentclass[conference]{IEEEtran}

\ifCLASSINFOpdf
  \usepackage[pdftex]{graphicx}
  \DeclareGraphicsExtensions{.pdf,.jpeg,.png}
\else
  \usepackage[dvips]{graphicx}
\fi
\usepackage{epstopdf}
\usepackage[cmex10]{amsmath}
\usepackage{amssymb}
\usepackage{wasysym}
\usepackage{algpseudocode}
\usepackage{algorithmicx,algorithm}
\usepackage{tabularx} 
\usepackage{array}
\usepackage{cite}
\usepackage{color}
\usepackage{url}
\usepackage{enumerate}
\usepackage{booktabs}
\usepackage{epsfig,latexsym}
\usepackage{flushend}
\usepackage{graphicx}
\usepackage{stfloats}
\usepackage{float}
\usepackage{bm}
\begin{document}
\title{Compressed Sensing based Detection Schemes for Differential Spatial Modulation in Visible Light Communication Systems}

\author{Zichun Shi$^{*}$, Pu Miao$^{*}$, Peng Chen$^{\ddagger}$, Lei Xue$^{\star}$, Li-Yang Zheng$^{\spadesuit}$, Laiyuan Wang$^{\dagger}$, Gaojie Chen$^{\dagger}$\\

$^*$School of Electronic and Information Engineering, Qingdao University, Qingdao, China\\
$^{\ddagger}$School of Information Sciences and Engineering, Southeast University, Nanjing, China\\
$^{\star}$School of Cyber Science and Technology, Sun Yat-sen University, Shenzhen, China\\
$^{\spadesuit}$School of Science, Sun Yat-sen University, Shenzhen, China\\
$^{\dagger}$School of Flexible Electronics, Sun Yat-sen University, Shenzhen, China\\
$^{*}$Email: mpvae@qdu.edu.cn\\
}

\maketitle

\begin{abstract}
Differential spatial modulation (DSM) exploits the time dimension to facilitate the differential modulation, which can perfectly avoid the challenge in acquiring of heavily entangled channel state information of visible light communication (VLC) system. However, it has huge search space and high complexity for large number of transmitters. In this paper, a novel vector correction (VC)-based orthogonal matching pursuit (OMP) detection algorithm is proposed to reduce the complexity, which exploits the sparsity and relativity of all transmitters, and then employs a novel correction criterion by correcting the index vectors of the error estimation for improving the demodulation performance. To overcome the local optimum dilemma in the atoms searching, an OMP-assisted genetic algorithm is also proposed to further improve the bit error rate (BER) performance of the VLC-DSM system. Simulation results demonstrate that the proposed schemes can significantly reduce the computational complexity at least by 62.5${\%}$ while achieving an excellent BER performance as compared with traditional maximum likelihood based receiver.
\end{abstract}

\section{Introduction}\label{sec1}
Light-emitting diode (LED) based visible light communication (VLC) is widely regarded as a promising technology for future high-speed indoor wireless communication due to its advantages such as low cost, license-free spectrum, no electromagnetic interference, and simultaneous illumination and communication \cite{ref-1,ref-3}. Nevertheless, the enhancement of the achievable spectral efficiency (SE) is impacted by the available bandwidth of LEDs \cite{ref-4}. Optical spatial modulation (SM) is kind of popular multiple-input multiple-output (MIMO) scheme and has attracted tremendous attention in band-limited VLC system, where the data bits are transmitted by selecting indexes of LEDs at each time for boosting the system capacity without increasing the power consumption \cite{ref-8}.

However, the original optical SM is coherent and the channel state information (CSI) must be known for an efficient decoding at the receiver \cite{ref-dsm}. Therefore, the introduced channel estimation error could result in performance loss inevitably. To this end, differential spatial modulation (DSM) which utilized a single transmit structure without any CSI was proposed \cite{ref-10}. In DSM, one out of LED matrices was activated to allocate symbols to the selected transmitters in different time instants to achieve the attainable high-data rate transmission\cite{ref-11,ref-111}. The maximum likelihood (ML) detector relies on a classical block-by-block based approach to jointly detect the activation transmitter matrix and the modulation symbols in a space–time block. Hence, the search space and computational complexity grow exponentially with the increasing of the set of symbol constellations and the number of transmitters. The results of \cite{ref-12} show that the likelihood function of the ML detection algorithm is independent among the rows thus the search space can be decomposed accordingly. In \cite{ref-13}, an approximate ML detector was proposed, where the index of the activated transmitters block in each time slot were obtained firstly. In \cite{ref-16}, a detection scheme based on the sparsity of the transmission matrix was proposed, where the index position of the largest element of the matrix was estimated after correlating the received symbol blocks of the previous and current time instants. A detection strategy based on inner products and infinite paradigms was proposed in \cite{ref-17}, which exploited the channel correlation during consecutive time slots to detect an activated transmitter, and then decoded the constellation symbols accordingly. Nevertheless, the reduction of computational complexity in above schemes is not be well achieved, which is one of the critical challenges in practical applications. 

In this paper, two novel detection schemes, in terms of vector correction (VC)-based orthogonal matching pursuit (OMP) and OMP-assisted genetic detection, are proposed by using the sparsity of transmitting matrix to reduce the detection complexity and also improve the BER performance of VLC-DSM. Simulation results show that the complexity of the two proposed algorithms can be reduced at least by 80${\%}$ and 62.5${\%}$ as compared to the traditional ML scheme, respectively. 

\emph{Notation:}
Let ${\left(  \cdot  \right)^T}$ denotes the transpose of a vector or matrix and $\left\lfloor  \cdot  \right\rfloor$ represents the floor operator, respectively. Let ${\left(  \cdot  \right)^{ - 1}}$ and ${\left(  \cdot  \right)^{H}}$ are used to represent the inverse and conjugate transpose of a matrix, respectively. In addition, $diag\left(\mathbf{a}\right)$ denotes a diagonal matrix with main diagonal vector $\mathbf{a}$, and ${\left\|  \cdot  \right\|_F}$ denotes the Frobenius norm operation.

\section{System Model}\label{sec2}

Assuming a general VLC-DSM system consisting of $N_t$ LEDs and $N_r$ PDs, where $M$-ary phase shift keying (PSK) is adopted. We select one index matrix ${{\mathbf{P}}_q }\in {\mathbb{R}^{{N_t} \times {N_t}}}\left( {q = 1,2,...Q} \right)$ from $Q$ dimensional activation LED index matrices to transmit $N_t$ constellation symbols. There is only one non-zero element per column in each activated LED index matrix. Therefore, the number of index matrices is ${N_t}!$ and each of them corresponds to an index vector ${{\mathbf{f}}_q}\left( {f_q^1,f_q^2,...,f_q^{{N_t}}} \right)$, where $f_q^i\left( {i = 1,2,...N_t} \right)$ denotes the position of the non-zero element in the $i$-th column of ${{\mathbf{P}}_q}$. However, in order to modulate the information bits, only $Q = {2^{\left\lfloor {{{\log }_2}\left( {{N_t}!} \right)} \right\rfloor }}$ index matrices are considered to be valid.

Assuming there are ${\rm T}$ transmission blocks in total, in each transmission block, the number of the input information bits is ${N_t}{\log _2}M + \left\lfloor {{{\log }_2}\left( {{N_t}!} \right)} \right\rfloor$, where the first $\left\lfloor {{{\log }_2}\left( {{N_t}!} \right)} \right\rfloor$ bits are encoded in the spatial domain in the form of the gray code, and the remaining ${N_t}{\log _2}M$ bits are used to modulate $N_t$ PSK symbols $\left[ {{p_1},{p_2},...,{p_{{N_t}}}} \right]$, then a unique bit information carrying matrix can be obtained as
\begin{equation}\label{Eq.1}
{{\mathbf{X}}_\tau } = {{\mathbf{P}}_q}diag\left[ {{p_1},{p_2},...,{p_{{N_t}}}} \right].
\end{equation}
where $\tau  \in \left\{ {0,1,...,{\rm T} - 1} \right\}$. Let ${{\mathbf{S}}_\tau } \in {\mathbb{R}^{{N_t} \times {N_t}}}$ denotes the signal transmission matrix of the $\tau$-th transmission block, where ${{\mathbf{S}}_0 }$ is the initial matrix, which is typically a unit matrix. Then, the differential relationship can be expressed as  
\begin{equation}\label{Eq.2}
{{\mathbf{S}}_\tau } = {{\mathbf{S}}_{\tau  - 1}}{{\mathbf{X}}_\tau }.
\end{equation}
Let ${{\mathbf{Y}}_\tau } \in {\mathbb{R}^{{N_r} \times {N_t}}}$ denote the received signal matrix of the $\tau$-th transmission block, and it can be expressed as
\begin{equation}\label{Eq.3}
{{\mathbf{Y}}_\tau } = {{\mathbf{H}}_\tau }{{\mathbf{S}}_\tau } + {{\mathbf{n}}_\tau },
\end{equation}
and
\begin{equation}\label{Eq.4}
{{\mathbf{Y}}_{\tau  - 1}} = {{\mathbf{H}}_{\tau  - 1}}{{\mathbf{S}}_{\tau  - 1}} + {{\mathbf{n}}_{\tau  - 1}}.
\end{equation}
By combining \eqref{Eq.2}, \eqref{Eq.3} and \eqref{Eq.4}, ${{\mathbf{Y}}_\tau }$ can be further simplified as
\begin{equation}\label{Eq.5}
{{\mathbf{Y}}_\tau } = {{\mathbf{Y}}_\tau }{{\mathbf{X}}_\tau } + {{\mathbf{n}}_\tau } - {{\mathbf{n}}_{\tau  - 1}}{{\mathbf{X}}_\tau },
\end{equation}
where ${{\mathbf{H}}_\tau } \in {\mathbb{R}^{{N_r} \times {N_t}}}$ denotes the VLC channel matrix, ${{\mathbf{n}}_\tau } \in {\mathbb{R}^{{N_r} \times {N_t}}}$ denotes additive Gaussian white noise (AGWN) with zero mean value. In addition, the channel matrix ${{\mathbf{H}}_\tau }$ can be given as
\begin{equation}\label{Eq.6}
{{\mathbf{H}}_\tau }=\begin{bmatrix}
	{h_{11}} & \cdots & {h_{1{N_t}}} \\
	 \vdots  &  \ddots  &  \vdots  \\
	{h_{{N_r}1}}  &  \cdots  & {h_{{N_r}{N_t}}} \\
 \end{bmatrix},
\end{equation}
 the channel gain ${h_{ji }}  \left( {j = 1,2,\ldots,{N_r}} \right)$ between the $i$-th LED and the $j$-th PD follows the Lambertian radiation pattern, and can be calculated by
\begin{equation}
		\label{Eq.7}
		{h_{ji }} = \frac{{\left( {w + 1} \right)\rho A}}{{2\pi d_o^2}}{\cos ^w}\left( {{\varphi _o}} \right){T_s}\left( {{\theta _o}} \right)g\left( {{\theta _o}} \right)\cos \left( {{\theta _o}} \right),
	\end{equation}
	where $w =  - \ln 2/\ln \left( {\cos \left( \Psi  \right)} \right)$ represents the Lambertian emission order, $\Psi $ is the semi-angle at half power of the LED. The responsivity is denoted by $\rho $, and $A$ stands for the PD’s active area. Additionally, ${d_o}$, ${\varphi _o}$, and ${\theta _o}$ refer to the distance, the angle of emission, and the angle of incidence between the $i$-th LED and the  $j$-th PD, respectively. ${T_s}\left( {{\theta _o}} \right)$ represents the gain of optical filter, $g\left( {{\theta _o}} \right) = \frac{{{n^2}}}{{{{\sin }^2}\Phi }}$ is the gain of optical lens, where $n$ is the refractive index and $\Phi $ is the half-angle field of view (FOV) of the optical lens.

At the receiver, the ML detection can be expressed as
\begin{equation}
		\label{Eq.8}
		{{\mathbf{S}}_\tau } = \mathop {\arg \min }\limits_{{{\mathbf{X}}_\tau } \in \mathbb{X}} \left\| {{{\mathbf{Y}}_\tau } - {{\mathbf{Y}}_{\tau  - 1}}{{\mathbf{X}}_\tau }} \right\|_F^2,
	\end{equation}
where $\mathbb{X}$ denotes the set of all possible $\mathbf{X}_\tau$. As can be seen from \eqref{Eq.8}, the ML detector needs to traverse all the matrices to get the possible results, leading to a large computational complexity consequently.

\section{Proposed Compressed Sensing-based detectors}\label{sec3}
\subsection{The Proposed VC-based OMP Detection}
As for VLC-DSM, the design of symbol detection should consider the differential coding rather than to directly estimate the transmit signal vector from the received signal vector and channel matrix at the current moment. Let ${\mathbf{y}}_\tau ^i$ denotes the received signal vector of the $i$-th slot in the $\tau$-th block, then the vector ${\mathbf{y}}_\tau ^i$ can be expressed as
\begin{equation}
		\label{Eq.9}
		{\mathbf{y}}_\tau ^i = {{\mathbf{Y}}_{\tau  - 1}}{\mathbf{x}}_\tau ^i + {\mathbf{\hat n}}_\tau ^i,
	\end{equation}
where ${\mathbf{y}}_\tau ^i$, ${\mathbf{x}}_\tau ^i$ and ${\mathbf{\hat n}}_\tau ^i$ denote the $i$-th column of ${{\mathbf{Y}}_\tau }$, ${{\mathbf{X}}_\tau }$ and ${{\mathbf{n}}_\tau } - {{\mathbf{n}}_{\tau  - 1}}{{\mathbf{X}}_\tau }$, respectively. Note, ${\mathbf{x}}_\tau ^i$ is a sparse matrix with a sparsity of 1. In order to detect the activation LED index vector of the $i$-th time slot in the $\tau$-th block, the ${{\mathbf{Y}}_\tau }$ and ${{\mathbf{Y}}_{\tau  - 1}}$ need to be firstly normalized as ${{\mathbf{\bar Y}}_\tau }$ and ${{\mathbf{\bar Y}}_{\tau  - 1}}$, respectively. Then, the measurement matrix can be expressed as 
\begin{equation}
		\label{Eq.10}
		{{\mathbf{Y}}_{nom}} = {{\mathbf{\bar Y}}_\tau }^H{{\mathbf{\bar Y}}_{\tau  - 1}}.
	\end{equation}
Therefore, the detection can be formulated by
\begin{equation}
		\label{Eq.11}
		{\mathbf{y}}_\tau ^i = {{\mathbf{Y}}_{nom}}{\mathbf{x}}_\tau ^i.
	\end{equation}
 
Let $e$ denote the number of iterations, ${{\bf{r}}_e}$,  ${B_e}$ and ${\Lambda _e}$ denote the residual, atomic set and index set in the $e$-th iteration, respectively. Then, the initialization items can be set by $e = 1$, ${{\mathbf{r}}_0} = {\mathbf{y}}_\tau ^i$, ${\text{ }}{\Lambda _0} = \emptyset$, and ${B_e}=\emptyset$. The support index can be obtained by computing the inner product, shown as
\begin{equation}
		\label{Eq.12}
	{\xi _e} = \arg \max \left( {\left| {{\mathbf{Y}}_{nom}^H{{\bf{r}}_{e = 1}}} \right|} \right).	
	\end{equation}
Putting the index of the inner products with the maximum absolute values into the set ${\Lambda _e}$, the index set can be updated as 
\begin{equation}
		\label{Eq.13}
		{\Lambda _e} = {\Lambda _{e - 1}} \cup \left\{ {{\xi _e}} \right\}.
	\end{equation}
Then, the atomic set can be collected by 
\begin{equation}
		\label{Eq.14}
		{B_e} = {{{\mathbf{Y}}_{nom}}}(\xi _e) \cup {B_{e-1}}. 
	\end{equation}
Therefore, the estimated signal can be obtained by
\begin{equation}
		\label{Eq.15}
		{\hat s_e} = {\left( {{{\left( {{B_e}} \right)}^T}*{B_e}} \right)^{ - 1}}*{\left( {{B_e}} \right)^T}*{\mathbf{y}}_\tau ^i,
	\end{equation}
and the residual is
\begin{equation}
		\label{Eq.16}
		{{\bf{r}}_e} = {\mathbf{y}}_\tau ^i - {B_e}{\hat s_e}.
	\end{equation}
Putting the estimated signal value into the set ${\Gamma _e}$, and the estimated signal set can be obtained as
        \begin{equation}
		\label{Eq.17}
		{\Gamma _e} = {\hat s_e} \cup {\Gamma _{e - 1}}.
	\end{equation}
After repeating the process \eqref{Eq.12} to \eqref{Eq.17}, the signal vector ${\hat {\mathbf{s}}_\tau^i}$ and index vector ${\hat {\mathbf{v}}_\tau^i}$ are estimated consequently.

However, during the estimation of the index vectors, it is easy to encounter a situation that the demodulated spatial signal bits are all zeros, which will lead to the generation of illegal index vectors. To this end, we correct the estimated index vectors along with the signal detection after the above iteration, where the correction criterion can be described as followings.

\textbf{Criterion A }

\begin{enumerate}

\item[(a)] Compare the index vector ${\hat {\mathbf{v}}_\tau^i}$ with each vector of the legal index matrix $\mathbf F \left( {\mathbf{f}}_1, {\mathbf{f}}_2,...,{\mathbf{f}}_Q \right)$, which consisting of $Q$ legal activated LED index vectors. Finding the one that identical to ${\hat {\mathbf{v}}_\tau^i}$, then ${\hat {\mathbf{v}}_\tau^i}$ will be directly used as the final estimated index vector for demapping.

\item[(b)] If there is no one that is identical to ${\hat {\mathbf{v}}_\tau^i}$, then, the vector with the most same elements as vector ${\hat {\mathbf{v}}_\tau^i}$ is selected from matrix $\mathbf F$ as the final estimated vector.

\item[(c)] If there are multiple vectors in the matrix $\mathbf F$ satisfying the case appearing in step (b), the vector satisfying the condition that occurs first in the search order is preferred as the final one.

 \end{enumerate}

\subsection{The Proposed OMP-assisted Genetic Detection}
Although the above procedure can select the optimal atoms in each iteration, its search process may be limited by the local optimums. Therefore, we further introduce the genetic algorithm into the above scheme to detect the constellation signals and then improve the BER performance.

In order to involve the genetic algorithm into the procedure of DSM symbol detection, the initial population needs to be determined at the first step. Let ${{\bar y}_\tau ^i}$ denotes the $i$-th column of ${{\mathbf{\bar Y}}_\tau }$, and ${{\bar y}_{\tau-1} ^u}\left( {u = 1,2,...,{N_t}} \right)$ is the $u$-th column of ${{\mathbf{\bar Y}}_{\tau-1} }$. The inner product matrix ${{\mathbf{Y}}_{ipm}}$ can be calculated by the inner product of each column of ${{\mathbf{\bar Y}}_\tau }$ and ${{\mathbf{\bar Y}}_{\tau  - 1}}$, shown as 
\begin{equation}
		\label{Eq.ipm}
{{\mathbf{Y}}_{ipm}}\left( {u,i} \right) = \left| {\left\langle {\bar y_\tau ^i,\bar y_{\tau  - 1}^u} \right\rangle } \right|.
	\end{equation}
Let ${\mathbf{y}}_{ipm}^i$ denote the $i$-th column of ${{\mathbf{Y}}_{ipm}}$, and the corresponding index vector is expressed by ${{\mathbf{l}}_i}$. Sort ${\mathbf{y}}_{ipm}^i$ in descending order to obtain a new vector $\left( {{\mathbf{y}}_{ipm}^i} \right)'$ , and the corresponding index vector becomes ${{\mathbf{l}}_i}'$. Execute the above operation on all columns of the matrix ${{\mathbf{Y}}_{ipm}}$ to obtain an index matrix $\mathbf{L}$ with rearranged column elements. Each row of $\mathbf{L}$ is a possible combination of activated LED indexes, and the corresponding rearranged inner product matrix ${{\mathbf{R}}_{ori}}$ will be used as the initial population for subsequent genetic algorithm operations.

Assuming the number of individuals is ${A_{pop}}$, which is numerically consistent with the possible number of activated LED index combinations. The constellation symbols at each time slot can be determined by 
\begin{equation}
		\label{Eq.18}
\tilde s_{{f_q^i}} = \mathop {\arg \min }\limits_{s \in \Xi } \sum\limits_{j = 1}^{{N_r}} {{{\left| {{y_{ji}}\left( \tau  \right) - {y_{jf_q^i}}\left( {\tau  - 1} \right)s} \right|}^2}},
	\end{equation}
where $\tilde s_{{f_q^i}}$ denotes the constellation symbol of the ${f_q^i}$-th row of permutation matrix ${{\mathbf{P}}_q}$, and $\Xi $ is the set of $M$-ary PSK modulation symbols. Due to the fact that the selection probability is calculated based on the  fitness function, which is used for comparison and sorting in genetic iteration. Therefore, the fitness function needs to be set as positive one. According to the constellation symbols obtained from \eqref{Eq.18}, the fitness of individuals in the population can be selected as
\begin{equation}
		\label{Eq.19}
m = \sum\limits_{i = 1}^{{N_t}} {\sum\limits_{j = 1}^{{N_r}} {{{\left| {{y_{ji}}\left( \tau  \right) - {y_{jf_q^i}}\left( {\tau  - 1} \right)\tilde s_{{f_q^i}}} \right|}^2}} }.
	\end{equation} 
Therefore, it can be inferred that the lower the fitness value, the closer the detected signal approximates the correct signal.

To generate new individuals through pairwise crossover and delivery them to the next generation, it is necessary to select locally optimal individuals from the initial population ${{\mathbf{R}}_{ori}}$ based on the fitness assessment of individuals. Since smaller individual fitness is better as shown in \eqref{Eq.19}, the tournament selection strategy is employed in our scheme. The specific operations are as follows:

\textbf{Strategy A }

\begin{enumerate}

\item[(a)] Determining the number of individuals per selection.

\item[(b)] Randomly selecting individuals with the same probability to form a group, and then picking up the individual with the lowest fitness to be delivered to the next generation.

\item[(c)] Repeat step (b) until the size of the new population reaches the total size of the original ones.
\end{enumerate}
Afterwards, a signal matrix crossover strategy can be also adopted to foster the individuals with lower fitness in the subsequent generation, shown as:

\textbf{Strategy B }

 \begin{enumerate}
\item[(a)] In each group, the original population ${{\mathbf{R}}_{ori}}$ with two individuals are grouped as parents, which are crossed with the parents of the other groups at the probability of ${p_c}$.
\item[(b)] Then, a crossover bit is chosen at random within each group's dimension, i.e., a pair of parents crosses to produce a pair of offspring, where the crossover position ranges from $2$ to $Q - 1$. 
 \end{enumerate}
It should be noted that the range of the probability ${p_c}$ is set suitable to avoid these unexpected situations such as falling into a local optimum or the reduction of search speed. 
As for other crossover probability, we also devise a replication strategy to optimize individuals, shown as:

\textbf{Strategy C }

\begin{enumerate}
\item[(a)] Compare the metrics of the individual parents.
\item[(b)] In each group, eliminate the individuals with high m values and replace them with low m value individuals. 
 \end{enumerate}
Moreover, variation is performed on each bit of the individual with probability of ${p_m}$ to avoid the population falling into a local optimum, which can be expressed as
\begin{equation}
		\label{Eq.20}
{{\mathbf{R}}_{ori}}\left( {n,i} \right) = \left\{ {\begin{array}{*{20}{l}}
  {{{\mathbf{R}}_{ori}}\left( {n,i} \right),{\text{ }}if(p_{rand} > {p_m})} \\ 
  {{d_{rand}},{\text{ }}if(p_{rand} < {p_m})} 
\end{array}}, \right.
	\end{equation} 
where ${{\mathbf{R}}_{ori}}\left( {n,i} \right)$ denotes the $i$-th bit of the $n$-th individual, ${d_{rand}}$ is a random number that takes the value 0 or 1. All of the above procedures need to be iterated for ${G_t}$ times, and then the new population ${{\mathbf{R}}_{new}}$ will be obtained accordingly. 

\subsection{Complexity Analysis}
For the iteration of the proposed VC-based OMP detector (abbreviated as \textit{Scheme A}), the complexity is mainly determined by the inner product and least squares computation. The computational complexity of the inner product operation is related to the number of LEDs, which can be approximated as $O\left( {{N_t}} \right)$. The computational complexity required for the least squares problem is about $O\left( {N_r^2} \right)$. Since the process of vector correction occurs outside the iteration, the corresponding complexity can be negligible. In addition, the modulation order should be also taken into account, where its complexity can be approximated as $O\left( {{M}} \right)$. Therefore, the overall complexity can be approximately expressed as $O\left( {{M}N_r^2{N_t}} \right)$.

For the proposed OMP-Assisted genetic detector (abbreviated as \textit{Scheme B}), the complexity is mainly determined by three factors. First, the complexity of normalising the matrix can be approximately shown by $O\left( {{N_r}{N_t}} \right)$. The second is the calculation of inner product, which can be approximated expressed as $O\left( {{N_r}} \right)$. Finally, the constellation symbols are estimated and the complexity can be shown by $O\left( {M{N_r}{N_t}} \right)$. Since there are a total of ${A_{pop}}$ individuals and the entire process takes ${G_t}$ iterations, the complexity of the whole procedure can be expressed as $O\left( {M{N_r}{N_t}{G_t}{A_{pop}}} \right)$ approximately.

\section{Simulation Results}\label{sec4}
Considering the simulation environment of a room with dimensions of 4m$\times$4m$\times$3m,  where the receiving plane is placed at a height of 0.85m above the floor. We set the number is $N_t=2$ or $N_t=4$, and each transmitter is furnished with a single LED and positioned within a square array, maintaining a spacing of 1m between each unit. Additionally, the receiver equipped with a single PD is also arranged as a square array, and the number is $N_r=2$ or $N_r=4$. Both the semi-angle at half-power of LED and the half-angle FOV of optical lens are configured by 60 degrees. It is assumed that the PD possess a responsivity of 0.53A/W and an active receiving area of 1cm$^{2}$. Note that the transmitted signal noise ratio (SNR) was adopted in this paper for fair comparison between different MIMO schemes. 
 
\begin{figure}[t]
	\centering
\includegraphics[height=62mm, width=77mm]{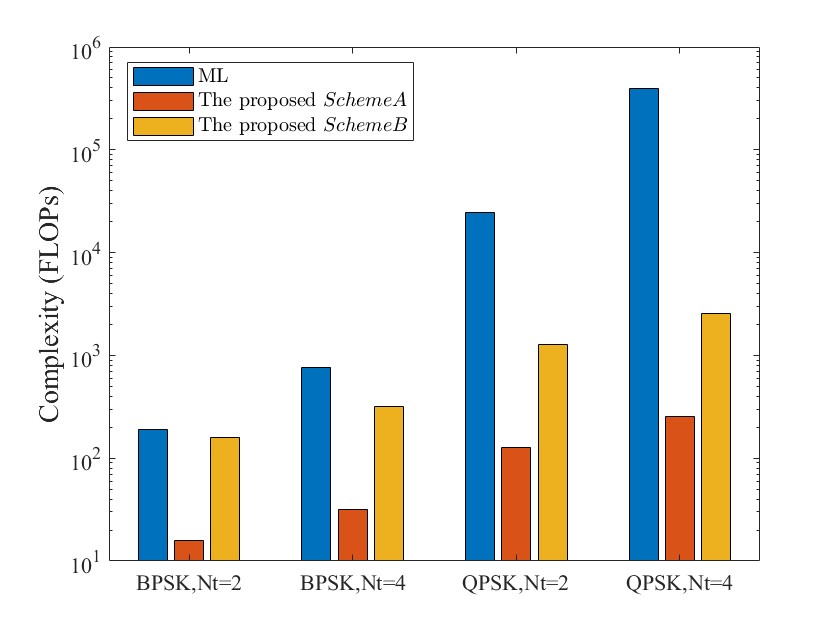}
	\caption{Comparison of computational complexity under different conditions.}
	\label{figure_2}
\end{figure}

 Fig.~\ref{figure_2} shows the complexity of each algorithm evaluated by using floating point operations under different conditions. The first set of histograms compares the complexity of the two proposed algorithms and the traditional ML algorithm for ${N_t} =2$ as binary phase shift keying (BPSK) is employed for simulation. It can be clearly seen that the complexity of the proposed \textit{Scheme A} is far reduced by about 80$\%$ compared to the ML algorithm, and the complexity of the proposed \textit{Scheme B} is only slightly lower than that of the ML algorithm. The second set of histograms compares the complexity of the three algorithms under BPSK modulation at ${N_t}=4$. The complexity of the proposed \textit{Scheme A} is still significantly lower than the other two schemes under this condition, and is reduced by 98$\%$ compared to the ML algorithm, while the complexity of the proposed \textit{Scheme B} is reduced by 62.5$\%$ compared to the ML algorithm, which is even more pronounced than that at ${N_t}= 2$. The third and fourth sets of histograms compare the complexity of the three algorithms under QPSK modulation, and the gap between the complexity of the proposed schemes under this condition does not change significantly due to the increase in the modulation order, whereas both proposed algorithms have a significant improvement in complexity compared to the ML algorithm, with at least a 99$\%$ reduction in the complexity of the proposed \textit{Scheme A}, and at least a reduction in the complexity of the proposed \textit{Scheme B} of about 91$\%$ compared to the ML algorithm. From the figure, it can be seen that as the number of LEDs and modulation order increases, the complexity of the ML algorithm exhibits an exponential growth, while the complexity of the two proposed schemes grows in the same order of magnitude, which is much lower than that of the ML, due to the fact that the amount of computation brought about by the iterative process of the two proposed algorithms does not increase with the change in the number of modulation order and the number of LEDs. 

\begin{figure}[t]
\centering
\includegraphics[height=62mm,width=77mm]{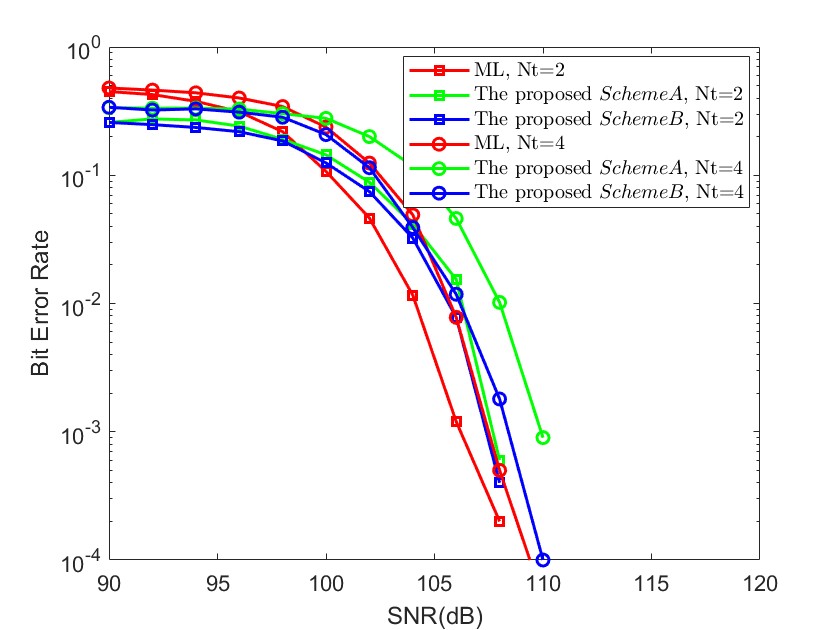}
\caption{BER performance comparison with BPSK constellation.}
\label{figure_3}
\end{figure}

\begin{figure}[t]
\centering
\includegraphics[height=62mm,width=77mm]{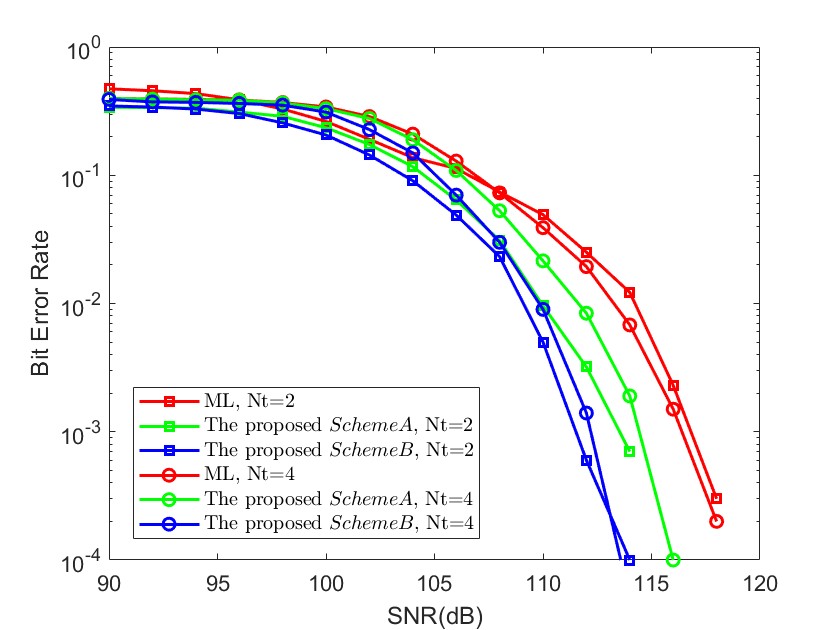}
\caption{BER performance comparison with QPSK constellation.}
\label{figure_4}
\end{figure}
 
Fig.~\ref{figure_3} compares the BER performance of the two proposed algorithms with the ML algorithm when BPSK modulation is adopted. The results clearly demonstrate that when the BER performance of the two proposed schemes are better than that of ML when the SNR is lower than 100 dB and same transmitters are employed. In addition, as the SNR moves more than 100 dB, the BER performances of the two proposed schemes are slightly inferior to the ML. The main reason is that there are only two constellation points for the set traversal in ML, which is favourable for achieving a low error probability with respect to brute force searching. However, as shown in Fig.~\ref{figure_2}, the ML achieved the affordable BER performance with much higher complexity than the two proposed schemes by sacrificing some computing resource. Based on the results of Fig.~\ref{figure_2} and Fig.~\ref{figure_3}, the proposed \textit{Scheme B} can achieve the best comprehensive performance, while the proposed \textit{Scheme A} shows a significant advantage in complexity. 

Fig.~\ref{figure_4} demonstrates the BER performance of the two proposed schemes when QPSK modulation is used under ${N_t} = 2$ and $4$, respectively. Overall, the BER performance of the two proposed schemes are better than that of the ML algorithm. Specifically, the BER performance of  \textit{Scheme A} can be improved by about 56$\%$ compared to the ML algorithm when the SNR is 110 dB, while that of \textit{Scheme B} will achieve 76.9$\%$ improvement as compared to ML detection. Therefore, both the \textit{Scheme A} and \textit{Scheme B} can achieve significant BER performance improvement with an excellent computational complexity advantage for QPSK constellation as compared with ML detection at the same condition.

\section{Conclusions}\label{sec5}
By reasonable exploiting the sparsity of transmitting matrix in VLC-DSM system, the VC-based OMP and OMP-assisted genetic detection schemes are proposed in this paper to reduce the computational complexity for DSM symbol detection. A special correction criterion is involved in the OMP iteration to address the illegal index vectors, and the hybrid genetic algorithm is also combined to improve the symbol estimation.  Simulation results show that the overall complexity and BER performance improvement can be achieved by the two proposed schemes as compared with the traditional ML algorithm for QPSK constellation.

\section*{Acknowledgment}
This research was partially funded by the Shandong Provincial Natural Science Foundation under grant ZR2023MF096, and by the National Natural Science Foundation of China under grant 61801257.

\AtEndDocument{\par\leavevmode}         

\end{document}